\documentclass[usenatbib]{mnras}
\usepackage{newtxtext,newtxmath,graphicx,subcaption, placeins}

\newcommand\codename{\textsc{Monk}}
\newcommand\taut{\tau_{\rm T}}
\newcommand\sigmat{\sigma_{\rm T}}
\newcommand\ergs{\rm erg~s^{-1}}
\newcommand\gs{\rm g~s^{-1}}
\newcommand\rg{\rm GM/c^2}
\newcommand\tobs{\theta_{\rm obs}}
\newcommand\rc{R_{\rm c}}

\usepackage[T1]{fontenc}

\DeclareRobustCommand{\VAN}[3]{#2}
\let\VANthebibliography\thebibliography
\def\thebibliography{\DeclareRobustCommand{\VAN}[3]{##3}\VANthebibliography}

\date{Accepted XXX. Received YYY; in original form ZZZ}
\title[BHXRB polarization]{Investigating the X-ray polarization of lamp-post coronae in BHXRBs}

\graphicspath{{./figures/}}

\begin{document}

\author[Wenda Zhang et al.]{Wenda Zhang$^1$\thanks{E-mail: wdzhang@nao.cas.cn},
Michal Dov\v ciak$^2$, Michal Bursa$^2$, Vladim\' ir Karas$^2$, Giorgio Matt$^3$ and Francesco Ursini$^3$\\
$^1$National Astronomical Observatories, Chinese Academy of Sciences,
20A Datun Road, Beijing 100101, China\\
$^2$Astronomical Institute, Czech Academy of Sciences, Bo\v cn\' i II 1401, CZ-141 00 Prague, Czech Republic\\
$^3$Dipartimento di Matematica e Fisica, Universit\` a degli Studi Roma Tre, via della Vasca Navale 84, 00146 Roma, Italy}

\label{firstpage}
\pagerange{\pageref{firstpage}--\pageref{lastpage}}
\maketitle

\begin{abstract}
High-sensitivity X-ray polarimetric observations of black hole X-ray binaries, which will soon become available
with the launches of space-borne X-ray observatories with sensitive X-ray polarimeters,
will be able to put independent constraints on the black hole as well as the accretion flow, and possibly break degeneracies 
that cannot be resolved by spectral/timing observations alone. 
In this work we perform a series of general relativistic Monte--Carlo radiative transfer simulations to
study the expected polarization properties of X-ray radiation emerging from lamp-post coronae in black hole X-ray binaries. 
We find that the polarization degree of the coronal emission of black hole X-ray binaries is sensitive to the 
spin of the black hole, the height of the corona, and the dynamics of the corona.
\end{abstract}


\section{Introduction}
Measuring the spin of black holes is an essential topic in contemporary astrophysics. For accreting stellar-mass black holes,
the black hole spin can be measured by several methods. One is the continuum-fitting method \citep{zhang_black_1997}, in which
the spin of stellar-mass black holes are constrained by fitting their continuum X-ray spectra.
This approach turns out to be particularly useful to study stellar-mass black holes, because their accretion discs produce
sufficiently clear signal of the thermal (multicolor) continuum with signatures of general relativity in soft X-rays
\citep{zhang_black_1997,li_multitemperature_2005}.
Another popular method is the iron line fitting method, 
where the spin is measured by fitting the relativistically broadened
iron K $\alpha$ line \citep[e.g.][]{reynolds_fluorescent_2003,miller_relativistic_2007}.
The latter approach has been applied with stellar-mass as well as supermassive black holes.
Both methods have been extensively used in numerous studies \citep[e.g.][]{gierlinski_application_2001,
davis_testing_2006,shafee_estimating_2006,mcclintock_spin_2006,liu_precise_2008,gou_determination_2009,gou_spin_2010,gou_extreme_2011,
blum_measuring_2009,miller_stellar-mass_2009,reis_determining_2009,miller_nustar_2013,parker_nustar_2015,el-batal_nustar_2016,
miller_nicer_2018,xu_reflection_2018,xu_hard_2018}.

For the black hole X-ray binary (BHXRB) GRO J1655$-$40, inconsistent measurements were obtained with these two methods \citep{shafee_estimating_2006,reis_determining_2009}.
Interestingly, for this particular source where low- and high-frequency quasi-periodic oscillations (QPOs) are simultaneously detected,
one can also measure its black hole spin by associating the frequencies of the two QPOs with the Lense-Thirring precission and the Keplerian
frequencies at the innermost stable circular orbit, respectively \citep{motta_precise_2014}. However, the black hole spin measured from QPOs
are consistent neither with continuum fitting nor line fitting results. 

X-ray polarimetric observations of accreting stellar-mass black holes will be
able to put independent constraints on the black hole spin. The strong gravitational field of the black hole
rotates the polarization vector of the emerging X-rays,
leading to stronger de-polarization effects at larger value of spin \citep{connors_polarization_1980}.
\citet{dovciak_thermal_2008} and \citet{li_inferring_2009} modelled the polarization signatures of accreting stellar-mass black holes in
the thermal-dominated states, and found the polarization degree to be sensitive to the black hole spin.
\citet{schnittman_x-ray_2009} investigated how the self-irradiation changes the polarization signatures.
\citet{abarr_polarization_2020} studied the polarization properties of the X-ray emission from warped accretion discs.
\citet{taverna_towards_2020,taverna_spectral_2021} studied these effects
by treating the absorption of the disc atmosphere more realistically.
Dedicated X-ray polarimetric observations will soon become possible with the launch of 
the \textit{Imaging X-ray Polarimetry Explorer} \citep[IXPE;][]{weisskopf_imaging_2016} and
the \textit{enhanced X-ray Timing and Polarimetry mission} \citep[eXTP;][]{zhang_enhanced_2019}.

In addition to the spin of the black holes, polarimetric observations of BHXRBs in the low-hard state have also the potential to
constrain the physical properties of the hot coronae. \citet{schnittman_x-ray_2010} studied the polarization properties of wedge and
spherical coronae (inside truncated discs), and found the polarization degree to be dependent on the geometry of the corona and 
parameters of the black holes.

It's also possible that the x-ray emission originates from aborted jets or the base of relativistic jets \citep[e.g.][]{ghisellini_aborted_2004,markoff_going_2005}, in which cases the coronae are located above the discs. In this paper we investigate the polarization properties of coronae by modelling the observed signal from different geometries using Monte--Carlo methods. The case of the coronal emission in Active Galactic Nuclei is explored in an accompanying paper \citep{ursini_prospects_2022}. The results of the X-ray polarization of relativistic jets in BHXRBs will be presented in a follow-up paper.

\section{Method}
\subsection{The Numerical Code} 
The Monte--Carlo simulations are carried out with \codename{} \citep{zhang_constraining_2019}, a Monte--Carlo radiative 
transfer code that includes all general relativistic effects. \codename{} samples seed photons 
and ray-traces the emitted photons along 
null geodesics in curved spacetime of the Kerr black hole. If the photon travels through a plasma medium,
\codename{} also takes into account various interactions between the medium and the 
photons, including the Compton scattering, the free-free absorption, and the synchrotron 
self-absorption. In this paper we consider Compton scattering only.
\textsc{Monk} utilises the ``superphoton'' method where each photon is assigned a statistical weight $w$. 
The weight $w$ has the physical meaning of the number of photons generated
per unit time as measured in a distant observer’s frame \citep[see][]{zhang_constraining_2019}.

In particular, \codename{} supports polarized radiative transfer in curved spacetimes, which enables 
us to study the polarization properties as well. In this paper we restrict ourselves
to the case of linear polarization, as the Compton scattering induces the linear polarization only
\citep{berestetskii_relativistic_1971,connors_polarization_1980}.
When only linear polarization is considered, the 
polarization degree is relativistically invariable, while the polarization angle is
connected with a complex constant of motion: the Walker-Penrose constant $\kappa_{\rm wp}$
\citep{walker_quadratic_1970,chandrasekhar_mathematical_1983}.
While propagating the photon we conserve its Walker-Penrose constant. If the photon gets scattered off an
electron, we sample the energy of the photon, re-evaluate the polarization degree
and angle, and update the value of the Walker-Penrose constant. We assume Klein-Nishina Compton scattering cross section.
A detailed description of how \codename{} handles the scattering process can be found in \citet{zhang_constraining_2019}.

In order to simulate the signal expected at a detector we collect photons arriving at infinity.
For each photon we have the following information: its statistical weight $w$, energy $E_\infty$,
polar angle $\theta_{\rm obs}$ (measured from the accretion disc symmetry axis, see Sec.~\ref{sec:model_assumptions}),
polarization degree $\delta$, and its polarization angle $\psi_{\infty}$ derived from $\kappa_{\rm wp}$.
The three Stokes parameters $I(E)$, $Q(E)$, and $U(E)$ (in units of energy per unit time per energy interval)
can be constructed with the following equations:
\begin{eqnarray}
I_E &=& \frac{4\pi\sum_k w_k}{\Delta E \Delta \Omega},\\
Q_E &=& \frac{4\pi\sum_k w_k\delta_k\ {\rm cos}2\psi_k}{\Delta E \Delta \Omega}, \\
U_E &=& \frac{4\pi\sum_k w_k\delta_k\ {\rm sin}2\psi_k}{\Delta E \Delta \Omega},
\end{eqnarray}
where the sum is performed over all photons with $\theta_{\infty} \in (i - \Delta i/2, i + \Delta i /2]$ and $E_{\infty} \in (E - \Delta E / 2, E + \Delta E / 2]$, $\Delta E$ is the width of the energy bin, $\Delta \Omega = 2\pi [{\rm cos}(i - \Delta i /2) - {\rm cos}(i + \Delta i /2)]$ is the solid angle subtended by the inclination bin, and $E_\infty$, $\theta_\infty$ are the energy and polar angle of the photons at infinity, respectively. In this work we take $\Delta i = 10^\circ$. As Compton scattering only induces linear polarization, we always maintain the Stokes parameter $V_E=0$. The corresponding polarization angle $\delta$ and the polarization degree $\psi$:
\begin{eqnarray}
 \delta &=& \frac{\sqrt{Q_E^2 + U_E^2}}{L_E},\\
 \psi &=& \frac{1}{2} {\rm atan}\frac{U_E}{Q_E}.
\end{eqnarray}
We define the polarization angle in such a way that the polarization angle is 0 or $\pi/2$
if the polarization vector is parallel with or perpendicular to
the plane defined by the line of sight and the accretion disc symmetry axis, respectively.

\subsection{Model Assumptions}
\label{sec:model_assumptions}
We assume a razor-thin Keplerian accretion disc located on the equatorial plane around a 
Schwarzschild or Kerr black hole. The disc extends down to the innermost stable circular orbit (ISCO). 
In the rest frame of the disc fluid, the emission of the disc follows a blackbody spectrum with an effective temperature 
that follows the Novikov-Thorne temperature profile 
\citep{novikov_astrophysics_1973,page_disk-accretion_1974}. We also adopt a color 
correction factor of 1.7. Throughout the paper we assume a black hole mass of $10~\rm M_\odot$.

The photons emerging from the disc midplane experience scattering before escaping the 
photosphere if the disc atmosphere is partially or fully ionized. Due to the
scattering process, the radiation becomes anisotropic and polarized.
We utilise analytical formulae of \citet{chandrasekhar_radiative_1960}, chapter X, to
evaluate the angular distribution and polarization degree of the disc photons.

\begin{figure}
\begin{subfigure}{.3\columnwidth}
\includegraphics[width=\textwidth]{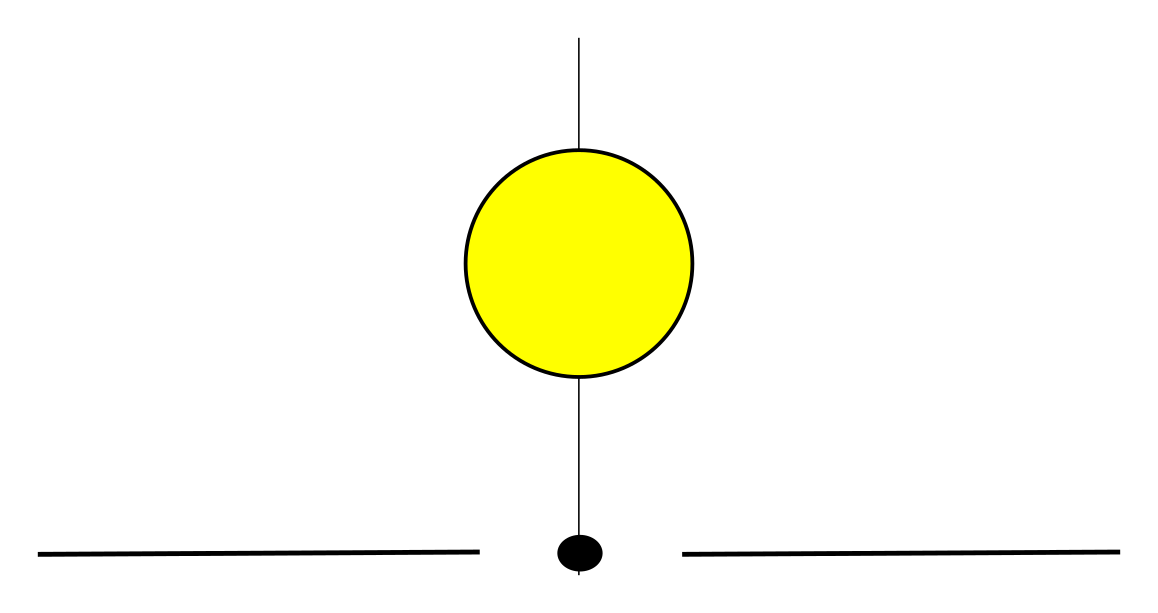}
\caption{Spherical coronae}
\end{subfigure}
\begin{subfigure}{.3\columnwidth}
\includegraphics[width=\textwidth]{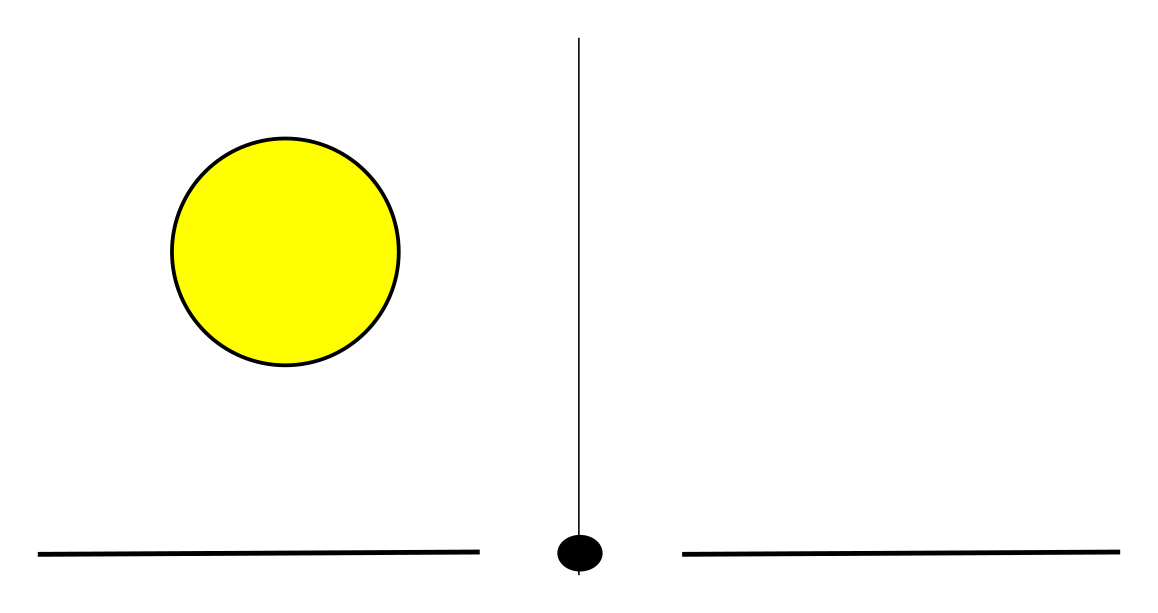}
\caption{Off-axis coronae}
\end{subfigure}
\begin{subfigure}{.3\columnwidth}
\includegraphics[width=\textwidth]{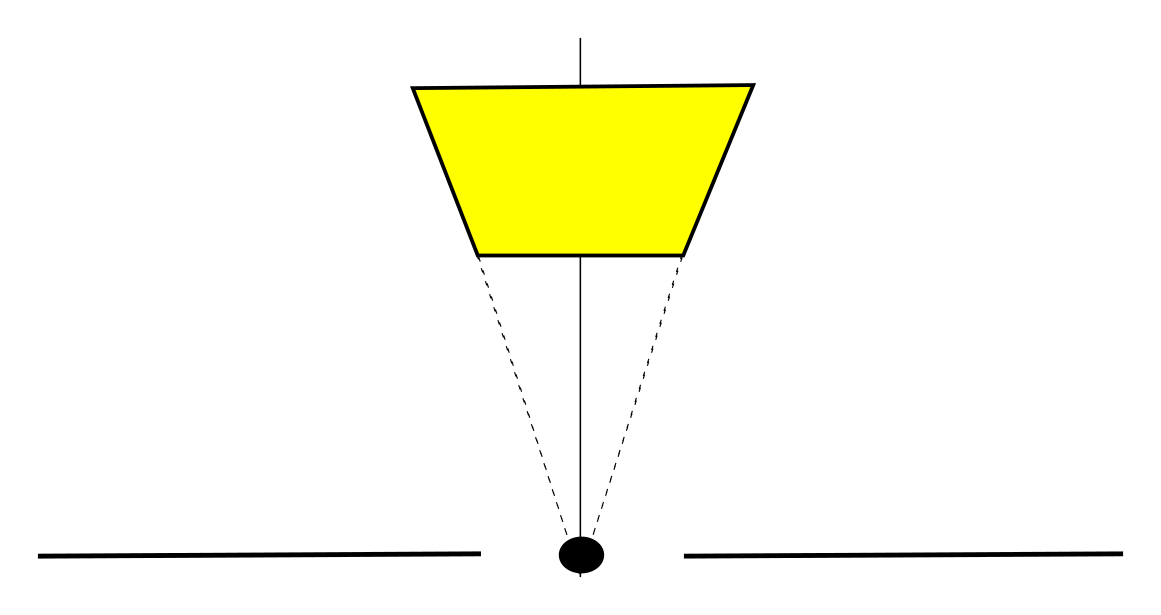}
\caption{Conical coronae}
\end{subfigure}
\caption{Schematic plots of the various corona geometries. In each subplot, the black circle represents the black hole horizon,
the horizontal line is a cross-section of a standard thin accretion disc, the vertical line represents the black hole rotation axis,
and the yellow region represents the corona.
\label{fig:geos}}
\end{figure}

We assume isothermal, homogeneous coronae. We assume thermal electrons whose velocity distribution follows the
Maxwell-J{\"u}ttner distribution. We consider three different geometries for the coronae:
\begin{itemize}

 \item Spherical coronae: illustrated in Fig.~\ref{fig:geos} (a). We assume that the spherical coronae are located on the black hole rotation axis. Motivated by \citet{beloborodov_plasma_1999}, we assume the coronae to be moving towards the $+z$ direction with velocity of $\beta=0.3$ as measured by a stationary observer. The coronae can be parameterized by the height above the disc $h$, the radius $R_c$, the electron temperature in the rest frame of the corona $kT_e$, and the Thompson optical depth $\taut \equiv n_e \sigmat R_c$, where $n_e$ is the electron density and $\sigmat$ is the Thompson scattering cross section.
 
 \item Off-axis spherical coronae: illustrated in Fig.~\ref{fig:geos} (b). Motivated by \citet{wilkins_understanding_2012}, we assume that the spherical coronae are above the disc, with its center off the black hole rotation axis. The coronae can be parameterized by the height above the disc $h$, the distance between the center of the corona to the black hole rotation axis $s$, the radius $R_c$, the electron temperature in the rest frame of the corona $kT_e$, and the Thompson optical depth $\taut \equiv n_e \sigmat R_c$. We assume the spherical coronae to be rotating about the black hole rotation axis with an angular velocity of $\omega = 1/(s^{3/2} + a)$, i.e. co-rotating with the underlying disc, where $a$ is the black hole spin.

 
 \item Conical coronae: illustrated in Fig.~\ref{fig:geos} (c). The corona can be parameterized by: the height of the corona base $h$, the thickness of the corona $\Delta h$, the opening angle $\alpha$, the Thompson optical depth of the corona $\taut \equiv n_e \sigmat \Delta h$, and the electron temperature in the rest frame of the corona $kT_e$. We let the Lorentzian factor of the bulk motion of the corona $\gamma=1/\Theta$ where $\Theta$ is the opening angle of the corona \citep[e.g.][]{blandford_relativistic_1979,daly_gasdynamics_1988,komissarov_magnetic_2009}. We assume that the corona moves upwards, towards the $+z$ direction.
 
\end{itemize}

\section{Results}

\subsection{Spherical coronae}
\label{sec:outspherical}
\begin{figure*}
\includegraphics[width=\textwidth]{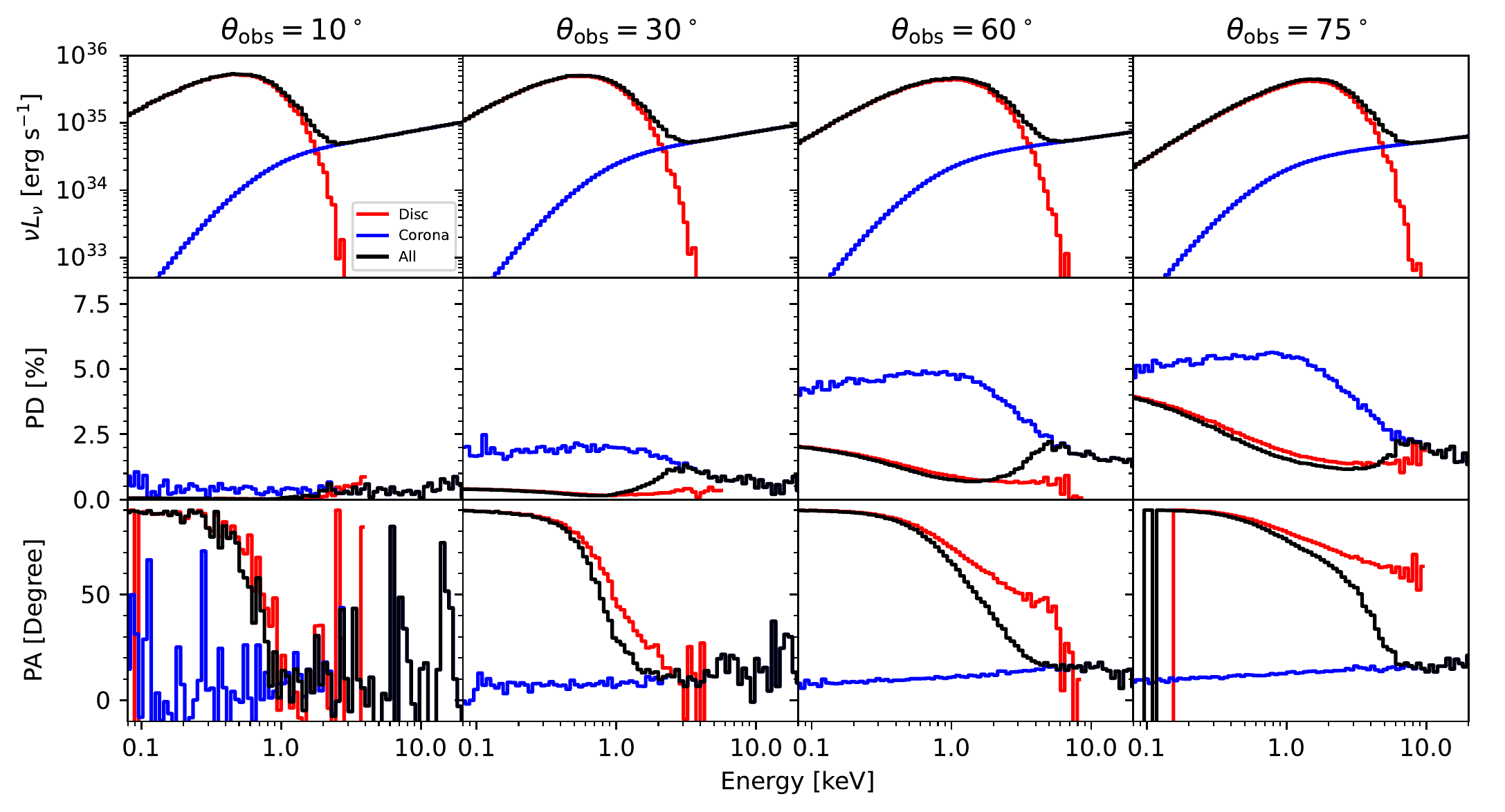}
\caption{From top to bottom: the energy spectra, polarization degree, and polarization angles of the radiation emerging from a disc-corona system 
around a stellar-mass black hole, as seen by distant observers at various inclinations.
The contributions of the disc and the coronae are plotted in red and blue colors, respectively, while the sum of the two are plotted in black.\label{fig:outsphere}}
\end{figure*}

In Fig.~\ref{fig:outsphere} we present the results for a disc-corona system that contains a 
spherical corona, for observers located at various inclinations. The parameters are: the
black hole spin $a=0.998$, the mass accretion rate $\dot{M}=4.32\times10^{15}~\rm g~s^{-1}$, 
$h=5~\rm \rg$, $R_c=3.5~\rm \rg$,
$\taut=1$, and $kT_e=100~\rm keV$. The energy spectrum (top panels) is dominated by 
thermal (red) and non-thermal (blue) radiation in the low- and high-energy regimes, respectively. 
The thermal radiation contains both seed photons that directly reach the observer and that pass through the corona without getting scattered.
The polarization degree (PD) of the thermal radiation decreases with energy, as high-energy thermal photons
originate from the disc close to the black hole where, the de-polarization effect of the
strong gravitational field of the black hole is most pronounced. 
The behavior of the thermal component is consistent with what was
found by \citet{dovciak_thermal_2008,li_inferring_2009} where the polarization
properties of the thermal radiation were thoroughly investigated.

The non-thermal radiation (blue) has a power-law shape with a photon index of $\sim 1.75$.
The PD of the non-thermal Comptonized radiation increases with energy, reaching a maximum
value at $\sim 1 ~\rm keV$, and then decreases with energy. The Comptonized emission is more polarized than
the thermal emission, and its PD increases with the observer's inclination,
from unpolarized at $\tobs=10^\circ$ to $\sim5\%$ at $\tobs=75^\circ$.
At all inclinations the PA of the Comptonized radiation is close to $0$.


\begin{figure}
 \includegraphics[width=\columnwidth]{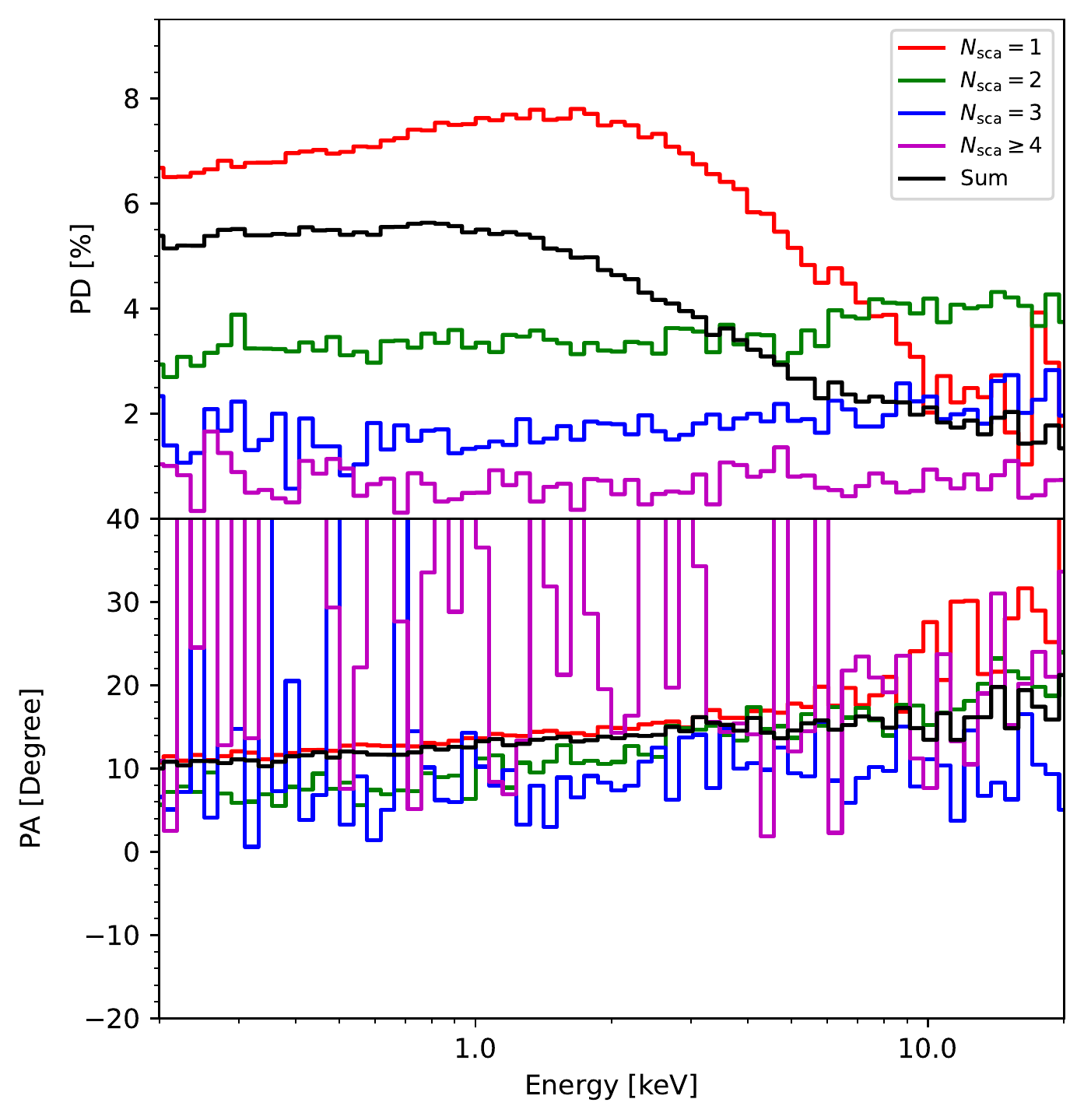}
 \caption{The PD (upper panel) and PA (lower panel) of the coronal emission, for various scattering orders
 in different colors. The observer is located at an inclination of
 75$^\circ$.\label{fig:outsphere_nsca}}
\end{figure}

To understand the behavior of the Comptonized emission, we extract the energy spectra and PD of different scattering orders. The result for $\tobs=75^\circ$ are presented in Fig.~\ref{fig:outsphere_nsca}. We find that the photons scattered once ($N_{\rm sca}=1$) are most polarized, with a peak PD of $\sim 8\%$ at $\sim 2~\rm keV$. The PD decreases with the scattering order, to 
$\sim3\%$ for $N_{\rm sca}=2$ and $\sim 1\%$ for $N_{\rm sca} \ge 3$, while the PAs are close to $0$ for all components. This explains why the PD of the Comptonized emission decreases beyond $\sim 1~\rm keV$ where the emission of higher scattering orders dominates. For the Comptonization process, the PD depends sensitively on the scattering angle, 
and the photons are most polarized if the scattering angle is $\pi/2$, 
i.e. the direction of the in-coming photon is perpendicular to the direction of the
scattered photon \citep[e.g.][]{berestetskii_relativistic_1971,connors_polarization_1980}.
Compared with the first-scattering photons where the in-coming photons originate from the disc,
the directions of the in-coming photons of high scattering orders are more isotropic, thus leading 
to lower polarization degree.


\subsubsection{Dependence on corona height and black hole spin}
\label{sec:outsphere_height}

\begin{figure}
 \includegraphics[width=\columnwidth]{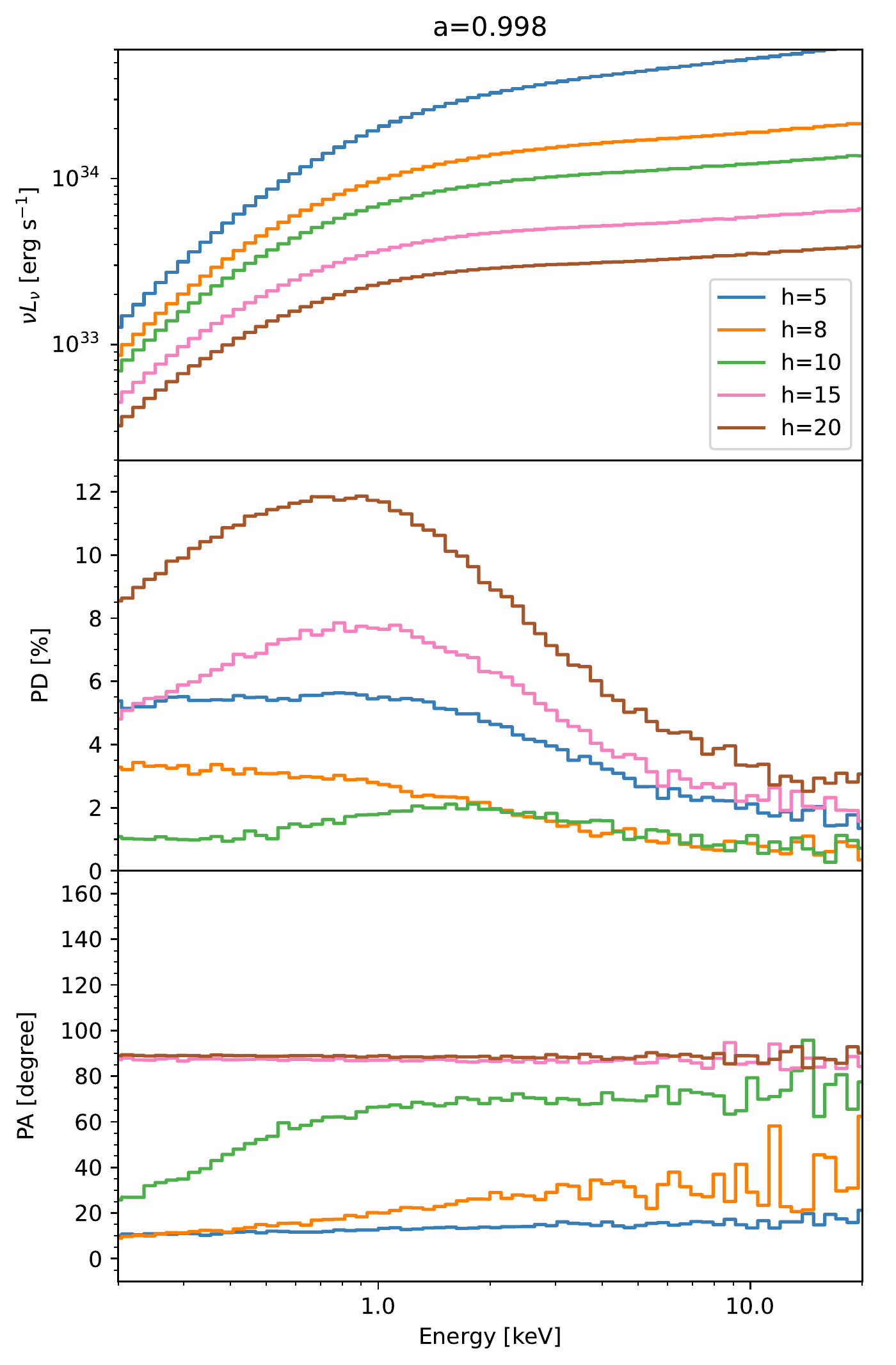}
 \caption{The energy spectra (top panel), PD (middle panel), and PA (bottom panel) of the coronal emission originating from spherical coronae. Results of different heights are plotted in different colors, as indicated in the plot. The observer is located at an inclination of $75^\circ$.
 \label{fig:outsphere_height_a1}}
\end{figure}

We also study how the polarization properties change with the height of the corona. In Fig.~\ref{fig:outsphere_height_a1} we compare the polarization properties of coronae with heights of $h= 5$ (blue), $8$ (orange), $10$ (green), $15$ (purple), and $20 ~\rg$ (brown) while keeping other parameters fixed as follows: $a=0.998$, $\dot{M} = 4.32\times10^{15}~\rm g~s^{-1}$, $\rc=3.5~\rg$, $kT_e=100~\rm keV$, and $\taut=1$. The observers are located at an inclination of $75^\circ$. The energy spectra (the upper panel) have similar shape. In contrast, the PD is quite sensitive to the height of the corona, with the maximum PD first decreasing from $\sim 5\%$ at $h=5~\rg$ to $\sim 3\%$ at $h=10~\rg$, then increasing to $\sim 11\%$ at $h=20~\rg$. The polarization angle also changes with the coronal height: when $h=5~\rg$, the PA is around 0. At intermediate heights ($h=10~\rg$), the PA increases from 0 to $\pi/2$ as the photon energy increase; while when $h=15$ and $20~\rg$, the PA is close to $\pi/2$.

Now we investigate the dependence of the polarization properties on the black hole spin. We perform simulations for Schwarzschild ($a=0$) black holes, assuming the same coronal heights with Kerr black holes (namely 5, 8, 10, 15, and 20$~\rg$). We set $\dot{M}=2.45\times 10^{16}~\gs$ Schwarzschild black holes, such that the power of the thermal emission is the same with Kerr black holes. For Schwarzschild black holes we set $\rc=2.5~\rg$, such that the corona at $h=5~\rg$ does not enter the black hole event horizon. Other properties are the same: $kT_e=100~\rm keV$ and $\taut=1$.

\begin{figure}
 \includegraphics[width=\columnwidth]{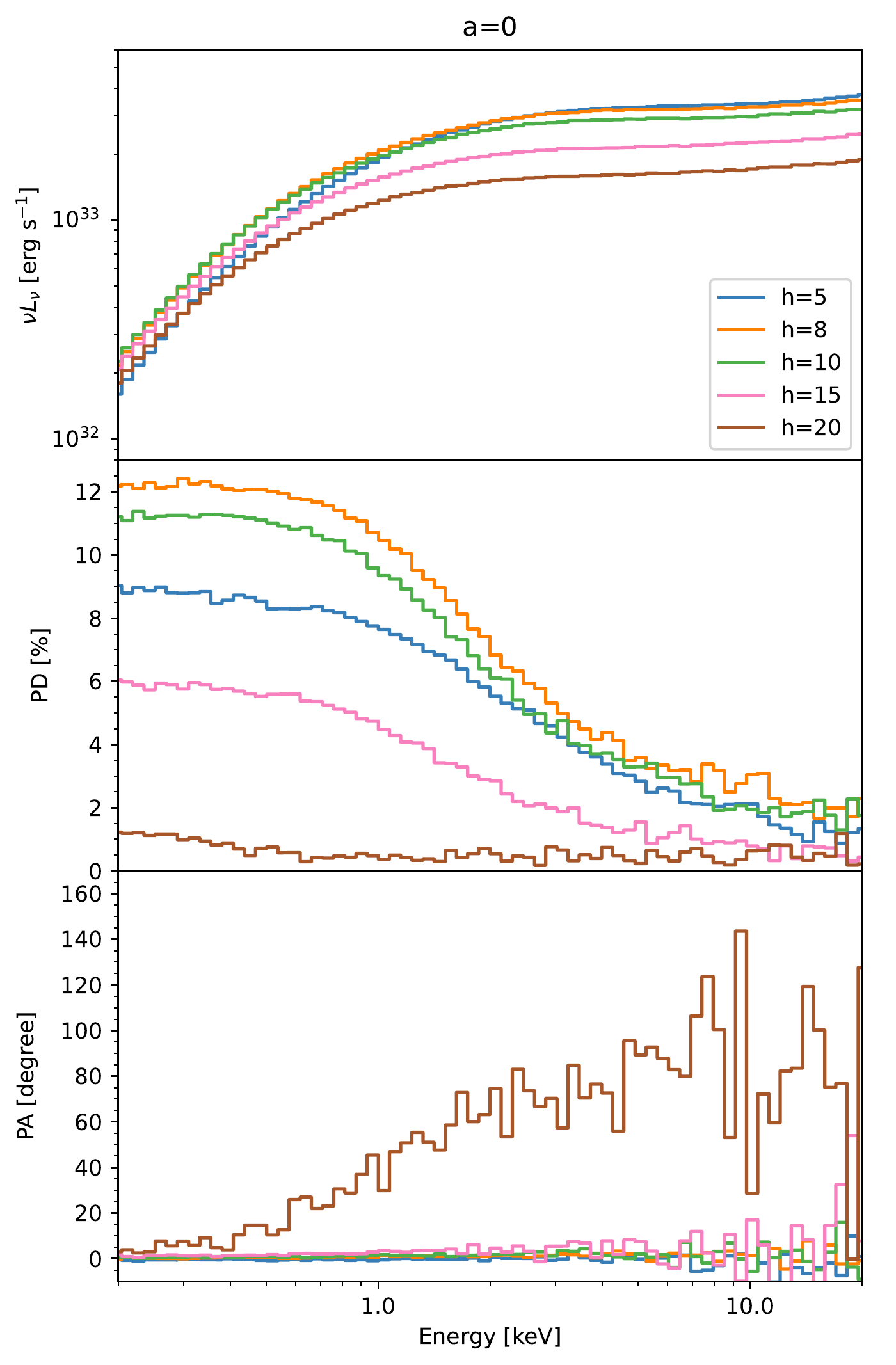}
 \caption{The same with Fig.~\ref{fig:outsphere_height_a1}, but for Schwarzschild black holes.
 \label{fig:outsphere_height_a0}}
\end{figure}

In Fig.~\ref{fig:outsphere_height_a0} we present the energy spectra, PD, and PA of the Componized emission for
Schwarzschild black holes, for observers located an inclination of $75^\circ$. While the energy spectra (the upper panel) have similar shape, the polarization properties are quite distinct. The PD is quite large at smaller heights ($\leq 10~\rg$), and decreases towards larger heights. The switch of the PA from 0 to $\pi/2$ is also seen, but at $h=20~\rg$, a larger height than Kerr black holes.

The reason for the dependence of the polarization properties on the coronal height and black hole spin is that, both parameters changes the incident angle of the seed photons that enters the corona. While for the height this is straightfoward, for black hole spin the change of the incident angle is due to the change in the ISCO. As we discussed earlier, the PD of the Comptonized radiation, in particular the first scattering component, is sensitive to the scattering angle.

\subsubsection{Dependence on mass accretion rate}
\begin{figure}
 \includegraphics[width=\columnwidth]{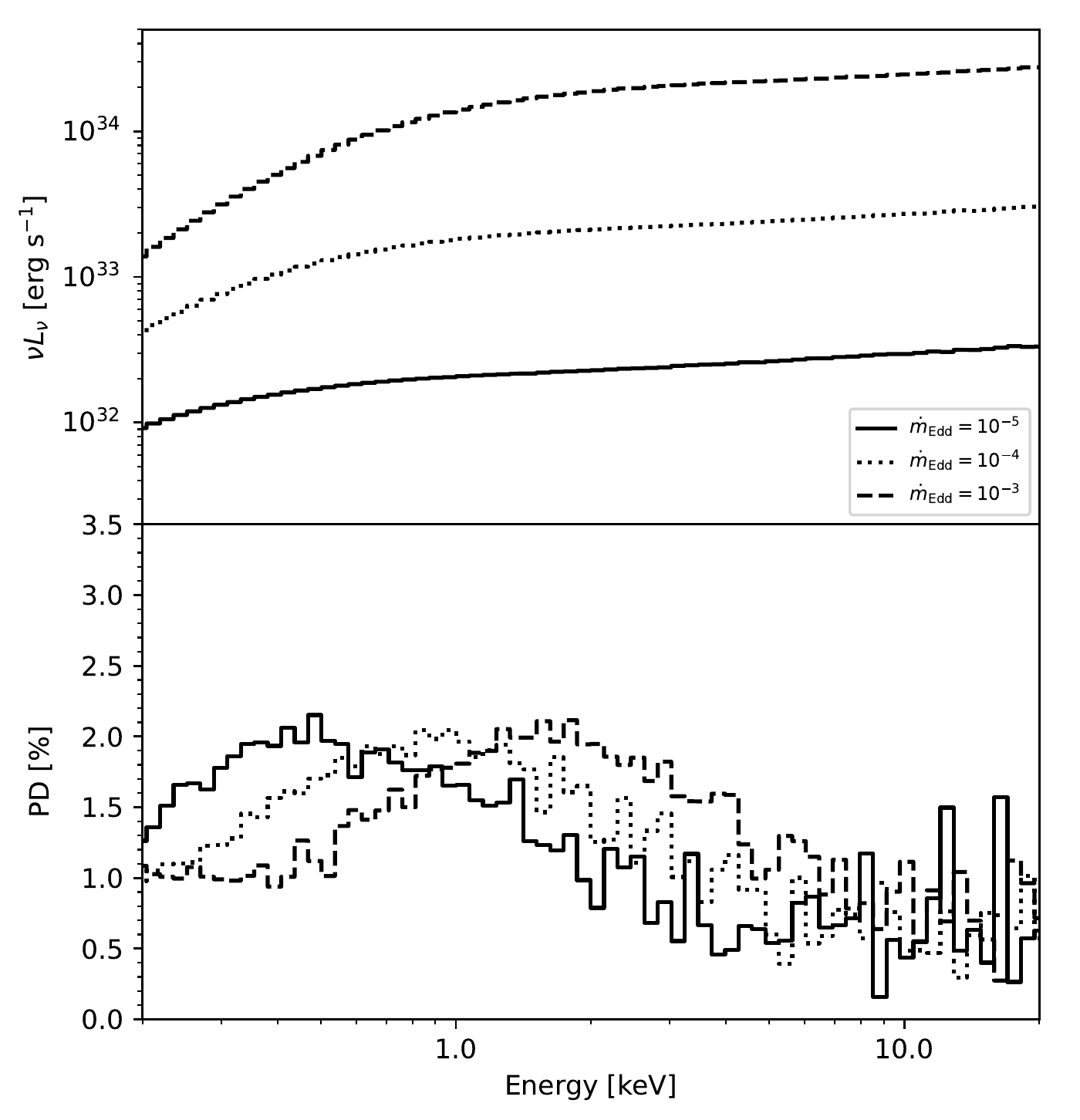}
 \caption{The energy spectra (upper panel) and PD (lower panel) of the coronal emission originating from spherical coronae. Results of different mass accretion rates are plotted in different line styles. The observer is located at an inclination of $75^\circ$.
 \label{fig:outsphere_mdot}}
\end{figure}

In this section we investigate the effect of changing $\dot{M}$, the black hole mass accretion rate. We pick three values of $\dot{M}$: $4.32\times10^{13}$, $4.32\times10^{14}$ and $4.32\times10^{15}~\rm g~s^{-1}$, such that the bolometric luminosity of the thin disc is $10^{-5}$, $10^{-4}$, and $10^{-3}$ Eddington luminosity, respectively. Other parameters are fixed: $a=0.998$, $h=10~\rg$, $\rc=3.5~\rg$, $kT_e=100~\rm keV$, and $\taut=1$. The observer is located at an inclination of $75^\circ$. In Fig.~\ref{fig:outsphere_mdot} we present the energy spectra (the upper panel) and PD (the lower panel) of the coronal emission for different $\dot{M}$. The peak PD remains more or less constant at $\sim 2\%$ as $\dot{M}$ changes, but the photon energy corresponding to the peak PD increases with $\dot{M}$, from $\sim0.4~\rm keV$ to $\sim1.1~\rm keV$. As shown in Fig.~\ref{fig:outsphere_nsca}, the photons that scattered once are most polarized. As $\dot{M}$ increases, the temperature of the seed photons also increases, leading to an increase of the peak energy of the photons that scatter once, and subsequently an increase of the energy corresponding to the peak PD.

\subsection{Off-axis coronae}

\begin{figure}
 \includegraphics[width=\columnwidth]{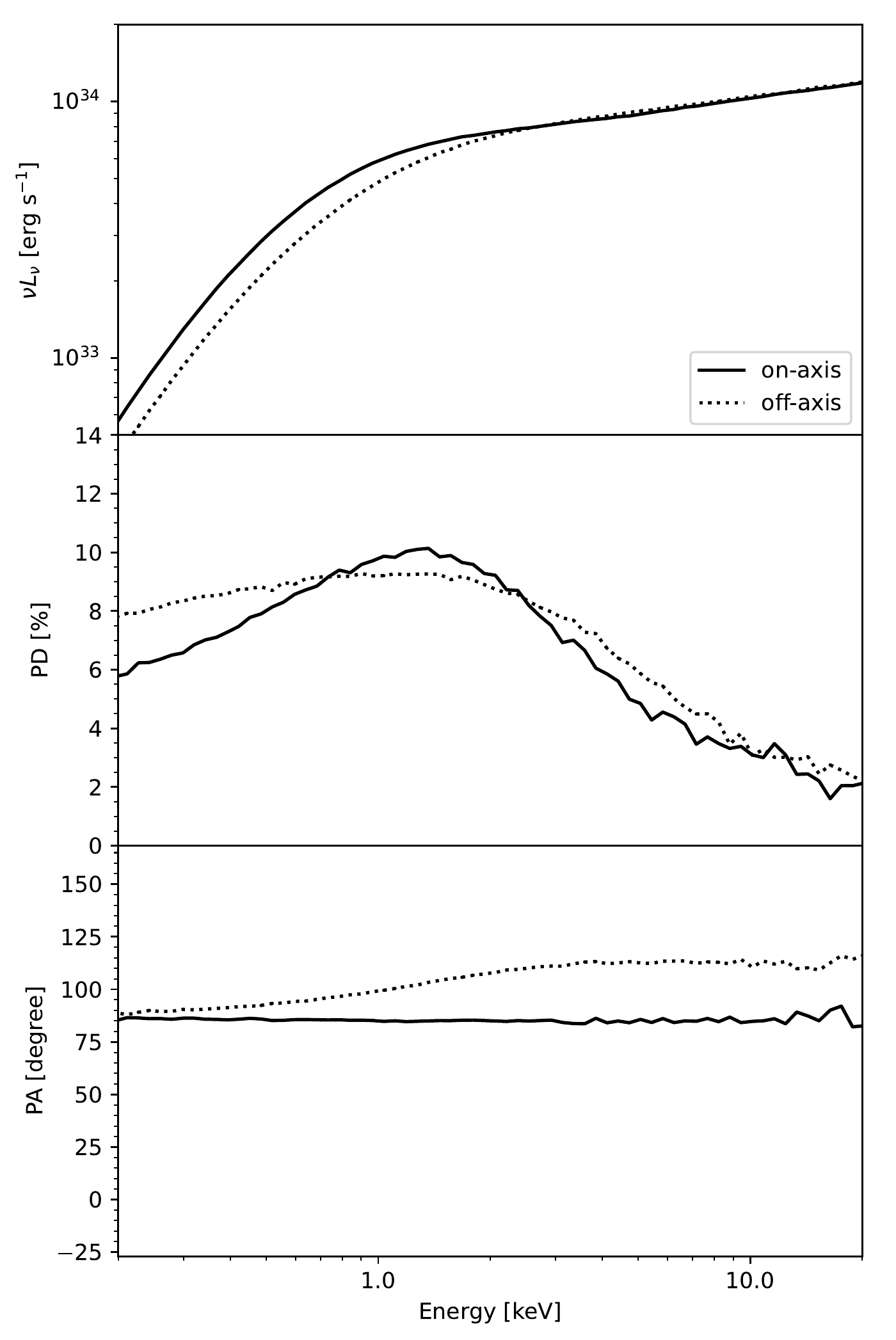}
 \caption{The energy spectra (upper panel), PD (middle panel), and PA (bottom panel) of the coronal emission originating from an on-axis (solid) corona and an off-axis corona (dotted). The observer is located at an inclination of $75^\circ$.
 \label{fig:rotsphere}}
\end{figure}

In Fig.~\ref{fig:rotsphere} we present the energy spectra, PD, and PA of the Comptonized emission for an off-axis corona with an offset of 10 $\rg$ from the symmetry axis, and compare the results with an on-axis coronae at the same heights. Other parameters are fixed: $a=0.998$, $\dot{M} = 4.32\times10^{15}~\rm g~s^{-1}$, $h=10~\rg$, $\rc=8~\rg$, $kT_e=100~\rm keV$, and $\taut=1$. The observer is located at an inclination of $75^\circ$. For the off-axis corona the emission is averaged over all orbital phases of the rotation, since the orbital period is short compared with any reasonable exposure time for X-ray polemetric observations. In both cases the energy spectra have photon indices of $\sim 1.8$ and luminosities $\sim 2\times10^{34}~\rm erg~s^{-1}$. The polarization properties are also similar, both with maximum PD of $\sim 10\%$ and PA of $\sim\pi /2$. This shows that the polarimetric properties do not change much between on- and off-axis coronae.

\subsection{Relativistic coronae}
\begin{figure}
 \includegraphics[width=\columnwidth]{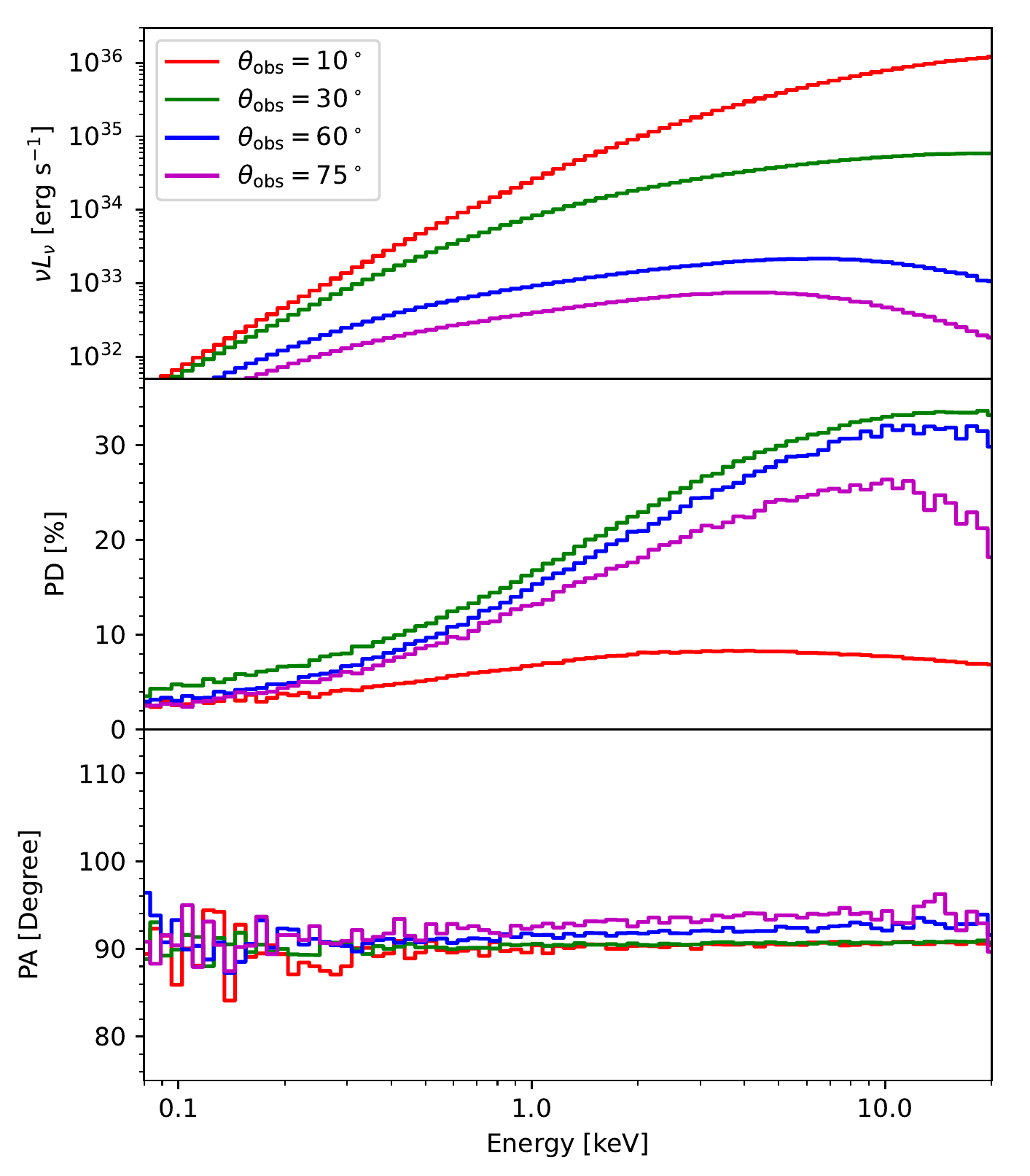}
 \caption{The energy spectra (top panel), PD (middle panel), and PA (bottom panel) of the coronal emission originating from a
 relativistic conical corona with $\gamma=4$. The results for different inclinations are plotted in different colors, as
 indicated in the plot.\label{fig:conical}}
\end{figure}

To assess the effect of the relativistic motion of the coronae on the polarization properties,
we simulate the emission from conical coronae moving along the
$+z$ direction with relativistic velocities (Fig.~\ref{fig:geos} (c)).
In Fig.~\ref{fig:conical} we present the energy spectrum, PD, and PA of the radiation from a disc-corona system containing a conical corona
with $kT_e=100~\rm keV$, $\taut=2$, and $\gamma=4$ (corresponding to an opening angle of $\sim 14^\circ$), for observers at various inclinations.
The emission is highly inclination-dependent: the $2-10~\rm keV$ luminosity of the
coronae emission decreases from $4.32\times 10^{35}~\ergs$ at $10^\circ$ to
$\sim 9.58\times 10^{32}~\ergs$ at $75^\circ$, due to stronger relativistic beaming effects at lower inclinations.
The spectral shape also varies with the observer's inclination.
We fit the $2-10~\rm keV$ spectra with a power-law model, and find the photon index to
be 0.63, 1.32, 1.71 and 1.99 for $\tobs=10^\circ$, $30^\circ$, $60^\circ$, and $75^\circ$, respectively.

The most striking feature of relativistic coronae is the high PD: when $\tobs\ge30^\circ$,
the PD can reach as high as $\sim 20-30\%$ above $\sim 3~\rm keV$,
much more polarized than stationary spherical/oblate coronae. 
The PD are also varying with $\tobs$: the peak PD is only $8\%$ when $\tobs=10^\circ$.
The PA of the coronal emission are always close to $90^\circ$.

\begin{figure}
 \includegraphics[width=\columnwidth]{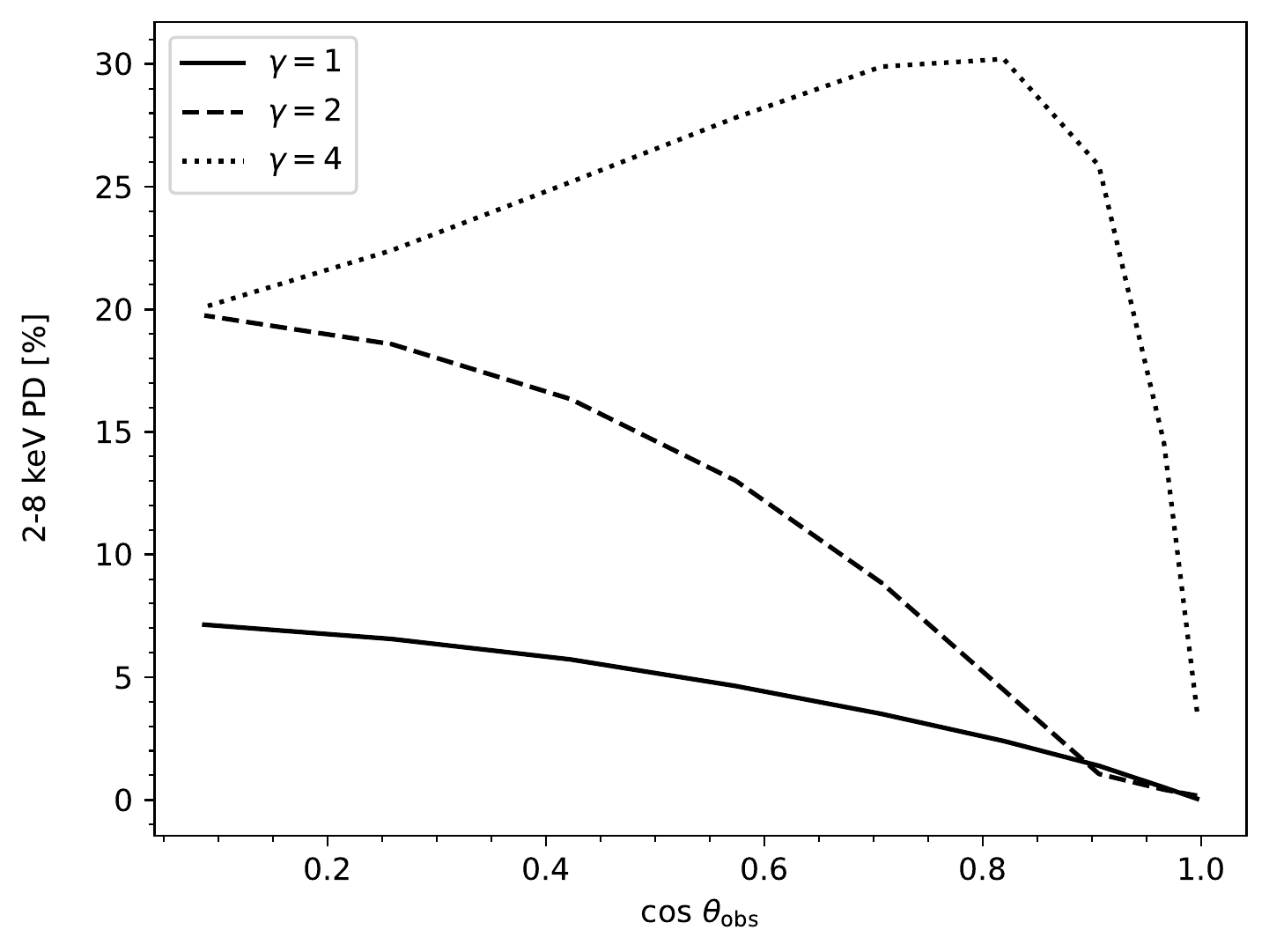}
 \caption{2--8 keV PD as a function of the observer's inclination, for conical coronae with various Lorentz factors.\label{fig:conical_gamma_incl}}
\end{figure}

In Fig.~\ref{fig:conical_gamma_incl} we compare the mean PD in the $2-8~\rm keV$ band (the energy band of \textit{IXPE})
of the coronal emission emitted by conical coronae with various Lorentz factors.
For comparison we also present the results for a stationary conical corona ($\gamma=1$).
The PD increases rapidly with $\gamma$: the maximum PD changes from $\sim 7\%$ to $\sim 30\%$ as $\gamma$ varies from $1$ to $4$.
While at low Lorentz factor ($\gamma=1, 2$), the PD decreases monotonically from an edge-on observer to a face-on observer, 
at high Lorentz factor ($\gamma=4$), the peak PD is reached at an inclination of $35^\circ$ and then decreases with the inclination.

\begin{figure}
 \includegraphics[width=\columnwidth]{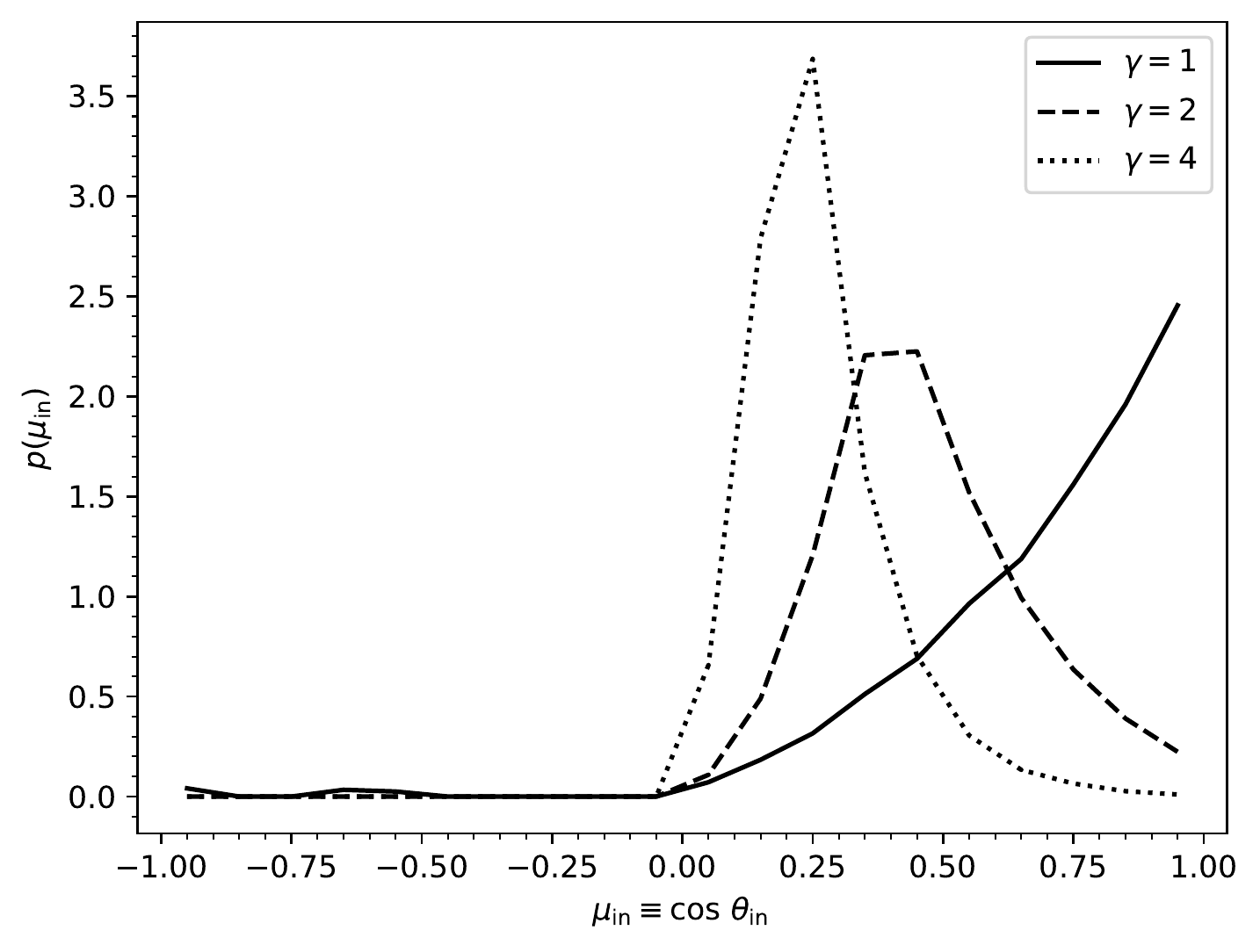}
 \caption{The probability distribution of ${\rm cos}\theta_{\rm in}$, where $\theta_{\rm in}$ is the incident angle of the seed
 photons as measured in the rest frame of the corona. Results of coronae with different velocities are plotted in different line
 styles, as indicated in the plot.\label{fig:scadir}}
\end{figure}

For coronae that move at relativistic velocity, the distribution of the direction of seed photons measured by fluids co-moving
with the corona gets much narrower as compared with stationary coronae. This can be seen in Fig.~\ref{fig:scadir} where we plot
$\mu_{\rm in} \equiv {\rm cos} \theta_{\rm in}$ as a function of the Lorentz factor $\gamma$ (here $\theta_{\rm in}$ is the angle between 
the seed photon and the black hole symmetry axis measured in the rest frame of the corona). For the corona with $\gamma=4$, the values
of $\mu_{\rm in}$ fall in a narrow range around 0.25, while for the stationary corona the distribution is much broader.
Since for the Compton scattering process the PD depends sensitively on the
scattering angle, we expect more polarized emission for narrower distribution of $\mu_{\rm in}$.
The distribution of $\mu_{\rm in}$ also explains the shift of the inclination corresponding to the peak PD with increasing $\gamma$,
as it is evident that the value of $\mu_{\rm in}$ corresponding to the peak of $\mu_{\rm in}$ distribution
changes with $\gamma$.

\section{Discussion}
\subsection{Spectral states}
\begin{figure}
 \includegraphics[width=\columnwidth]{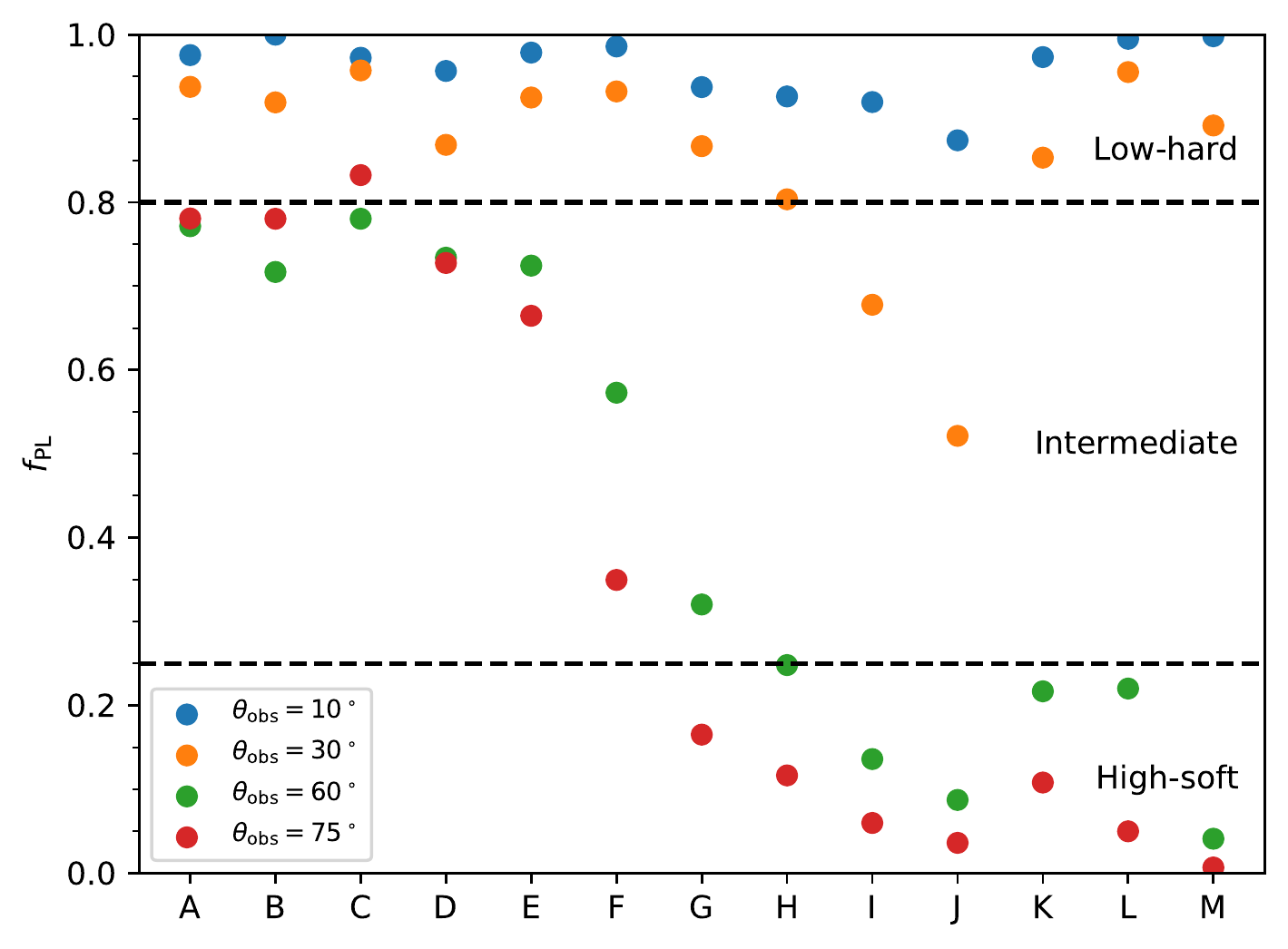}
 \caption{$2-20~\rm keV$ powerlaw fraction of various cases. 
 A-E: spherical coronae above Schwarzschild black holes, with heights of 5, 8, 10, 15, and 20~$\rg$, respectively.
 F-J: spherical coronae above extreme Kerr black holes, with heights of 5, 8, 10, 15, and 20~$\rg$, respectively.
 K: off-axis coronae. L: the conical corona with $\gamma=2$. M: the conical corona with $\gamma=4$.
 The results for different observer's inclinations are plotted in different colors, as indicated in the plot.\label{fig:state}}
\end{figure}

BHXRBs exhibit different accretion states in which the spectral and timing properties of their X-ray emission are quite distinct. While several definitions of the accretion states have been proposed, in this work we follow the definition of \citet{remillard_x-ray_2006}: the source is classified as high-soft (low-hard) state if the $2-20~\rm keV$ band powerlaw fraction $f_{\rm PL} < 25\%$ ($> 80\%$), and as intermediate state if otherwise. Note that the definition of \citet{remillard_x-ray_2006} makes use of timing information as well (such as the fractional rms), which is not available in our work. The accretion states corresponding to different cases in our work are presented in Fig.~\ref{fig:state}. It seems that the cases in our work cover all three spectral states, including the high-soft state. Among the cases classified as in the high-soft state, four of them are spherical coronae at large heights ($h \ge 10 ~\rg$; cases G-K) viewed at large inclinations. For $h \ge 8~\rg$, to compare with the case of smaller heights, we fix the coronal radius at a small value of $\rc = 3.5~\rg$, leading to a low powerlaw flux and subsequently a small value of $f_{\rm PL}$. Since the PD is not expected to change much with $\rc$, our conclusions can be applied to coronae with larger radii that have larger $f_{\rm PL}$ and corresponds to BHXRBs in the hard and intermediate states. Relativistic coronae (cases H \& I) viewed at high inclinations are also classified as in the high-soft state, since in this cases the coronal emission is beamed. However, it is known that relativistic jets are quenched in the high-soft state \citep{fender_towards_2004}. This indicates that the results of relativistic coronae viewed at high inclinations hardly correspond to any observation. Most other cases are classified as in the hard or intermediate states. Therefore our results are applicable to BHXRBs in the hard and intermediate states.

\subsection{Comparison with AGNs}
We find that the PD of spherical coronal in BHXRBs in the X-ray band can be as high as $\sim 12\%$. This is much larger than the PD of spherical coronae in AGNs, where a PD of only $1-3\%$ is expected \citep{ursini_prospects_2022}. This is not surpurising though, since for BHXRBs the seed photons have much larger temperature, and the first scattering photons, which is most polarized, can already reach the X-ray band. On the other hand, the X-ray emission of AGNs are contributed by photons that already scattered a few times, for which lower PD is expected.

%


\section{Summary}
In this work we investigate the polarization properties of lamp-post coronae around stellar-mass black holes. We find that the polarization property is sensitive to the properties of the black hole and the coronae. In particular:

\begin{itemize}
 \item The polarization degree of spherical, lamp-post coronae increases with the observer's inclination and decreases with the photon energy.
 \item The polarization properties of spherical, lamp-post coronae are sensitive to the black hole spin and the coronal heights. As the coronal height increases, the PD decreases and increases below and above a critical height, respectively. The PA switches from $0$ to $\pi/2$ at this critical height. The critical height increases with the black hole spin.
 \item The polarization properties of spherical, lamp-post coronae do not change much if the coronae are on or off the black hole rotation axis.
 \item The PD of coronal emission with different mass accretion rates have similar profiles, while the photon energy corresponding to the maximum PD increases with the mass accretion rate.
 \item The energy spectra and PD are very sensitive to the Lorentz factor of the coronae. The PD of the corona emission is boosted when the corona moves at relativistic velocity: at a Lorentz factor of 4, the PD of the coronal emission increase with the photon energy in the X-ray band, and can reach as high as $\sim 30\%$.
\end{itemize}

These results indicate that X-ray polarimetric observations of accreting stellar-mass black holes can offer us independent information of the systems that can be used to put constraints on the black hole spin, the coronal geometry, and the dynamics of the coronae. High-sensitivity X-ray polarimetric observations have become possible with the launch of the \textit{IXPE} in 2021 and \textit{eXTP} is planned to be launched in 2027. Our results are applicable to BHXRBs in the hard and intermediate states, where the Comptonized emission makes a significant contribution to the 2--8 keV band.


\FloatBarrier
\section*{Acknowledgements}
We thank the anonymous referee for his/her useful comments.
WZ acknowledges the support by the Strategic Pioneer Program on Space Science,
Chinese Academy of Sciences through grant XDA15052100.
MD and MB thank for the support from the GACR project 21-06825X and the institutional support from RVO:67985815.
VK acknowledges the Czech-Chinese Inter-Excellence mobility program (8JCH1080).
GM and FU acknowledge financial support from the Italian Space Agency (grant 2017-12-H.0).
This work is also supported by The Ministry of Education, Youth and Sports, Czech Republic, 
from the Large Infrastructures for Research, Experimental Development and Innovations project ``e-Infrastructure CZ - LM2018140''.
This research makes use of \textsc{matplotlib} \citep{hunter_matplotlib:_2007},
a Python 2D plotting library which produces publication quality figures.

\section*{Data availability}
The authors agree to make simulation data supporting the results in this paper available upon reasonable request.

%

\label{lastpage}
\end{document}